\documentclass[aps,pre,reprint,superscriptaddress,nofootinbib]{revtex4-1}
\usepackage{graphicx}
\usepackage{amsmath}
\usepackage{amsfonts}
\usepackage{hyperref}
\begin{document}

\title{Disordered surface vibrations in jammed sphere packings}

\author{Daniel M. Sussman}
\email[]{dsussman@sas.upenn.edu}
\author{Carl P. Goodrich}
\email[]{cpgoodri@sas.upenn.edu}
\author{Andrea J. Liu}
\email[]{ajliu@physics.upenn.edu}
\affiliation{Department of Physics and Astronomy, University of Pennsylvania, 209 South 33rd Street, Philadelphia, Pennsylvania 19104, USA}

\author{Sidney R. Nagel}
\email[]{srnagel@uchicago.edu}
\affiliation{The James Frank Institute, The University of Chicago - Chicago, Illinois 60637, USA }

\date{\today}

\begin{abstract}
We study the vibrational properties near a free surface of disordered spring networks derived from jammed sphere packings. In bulk systems, without surfaces, it is well understood that such systems have a plateau in the density of vibrational modes extending down to a frequency scale $\omega^*$. This frequency is controlled by $\Delta Z = \langle Z \rangle - 2d$, the difference between the average coordination of the spheres and twice the spatial dimension, $d$, of the system, which vanishes at the jamming transition. In the presence of a free surface we find that there is a density of disordered vibrational modes associated with the surface that extends far below $\omega^*$. The total number of these low-frequency surface modes is controlled by $\Delta Z$, and the profile of their decay into the bulk has two characteristic length scales, which diverge as $\Delta Z^{-1/2}$ and $\Delta Z^{-1}$ as the jamming transition is approached.
\end{abstract}

\maketitle

\section{Introduction}

Amorphous solids with free surfaces share a number of intriguing features. Nanometrically thin films of polymers and small-molecule glasses have glass-transition temperatures that are substantially lower than in bulk materials; nanoparticles display an excess of low-frequency modes in their vibrational densities of states compared to their bulk counterparts \cite{Sopu2011}; and free surfaces in nanopillars mediate the allowed failure modes that lead to shear banding \cite{Shavit2014, Greer2013}.  These findings are all correlated with the observation that relaxation dynamics are more rapid near a free surface than in the bulk~\cite{Ediger2014}.  The enhanced dynamics extend some distance into the bulk, but fail to correlate with measures of static structure that have been explored~\cite{Ediger2014}. An outstanding challenge is to find a structural feature that decays slowly enough from the surface that may be used to explain the increase in dynamics. More generally, the characteristic length scale over which a disordered solid is influenced by a free surface is unknown.

It is well-established in supercooled liquids that regions with large root-mean-squared short-time particle fluctuations are also regions that on longer time scales are more likely to exhibit particle rearrangements \cite{Harrowell2006}. Furthermore, these short-time fluctuations are themselves correlated with low-frequency, quasi-localized modes (which have low energy barriers to rearrangements \cite{Xu2010}) in both supercooled fluids \cite{Reichman2008,Royall2014} and jammed systems \cite{Manning2011}. The successful use of low-frequency modes to identify a structural population of potential flow defects in bulk systems \cite{Manning2011, Schoenholz2014} leads us to investigate the vibrational modes at the surface of model disordered systems. Specifically, we study disordered spring networks in dimensions $d=2$ and $d=3$. The networks are derived from jammed packings of soft spheres described in more detail below.  In the bulk, these networks are characterized by the average coordination of each particle, $\langle Z \rangle$, where the jamming transition occurs at the isostatic point where $\langle Z \rangle=Z_c=2d$ \cite{Ohern2003}. 

In bulk jammed systems a population of disordered low-frequency ``anomalous'' modes \cite{Wyart2005} swamp out the plane waves predicted by continuum elasticity. These additional modes can be understood as a consequence of a diverging length scale: as the jamming transition is approached from high density there is a diverging length scale $l^*\sim \Delta Z^{-1}$ where $\Delta Z \equiv (\langle Z \rangle-Z_c)$ \cite{Silbert2005,Wyart2005} that controls the effect of free surfaces on the stability of the system \cite{Wyart2005, Goodrich2013}. The low-frequency sound modes are connected to the zero-energy modes associated with uniform translations of the system, and similarly the anomalous modes are connected to zero-energy deformation modes that exist at the jamming transition in a system with free boundaries, according to a variational argument \cite{Wyart2005}.

Just as for systems with periodic boundary conditions, in disordered systems with a free surface the diverging length scales of jamming herald a new class of modes, and we find a robust population of disordered low-frequency vibrational modes localized near the surface. While there are zero-energy modes localized at the surface on the scale of a particle diameter \cite{Goodrich2013}, we find that the nonzero-frequency vibrations have an intricate spatial structure that extends into the bulk with length scales set by the proximity to the jamming transition. In addition to $l^*$ there is a second diverging length that controls system stability with respect to finite-wavevector boundary deformations~\cite{Silbert2005, Schoenholz2013}, the transverse length scale $l_T \sim \Delta Z^{-1/2}$. We find that this length is also relevant to disordered surface modes. These lengths, and other diverging lengths with the same scalings \cite{Ellenbroek2006, Lerner2014}, have been argued to characterize the length below which continuum elasticity fails \cite{Ohern2003} and the detailed disordered structure must be considered to understand the response of the amorphous material to point forces. 

The remainder of the paper is organized as follows. We begin in Sec. \ref{sec:prep} by describing the numerical preparation protocol for our systems. Section \ref{sec:harmresults} presents our numerical results on disordered spring networks, beginning in Sec. \ref{sec:dos} with data on the vibrational density of states and continuing in Sec. \ref{sec:struct} in which we investigate the spatial structure of the surface vibrational modes. We close with a discussion of these results in the context of the broader class of amorphous solids in Sec. \ref{sec:bridge}.

\section{System preparation \label{sec:prep}}

We begin by numerically generating jammed packings of $N$ bidisperse spheres in two and three dimensions.  We use two distributions (i) 50-50 mixture of spheres with diameter ratio 1:1.4 and (ii) a polydisperse mixture using a flat distribution of particle sizes between $\sigma$ and $1.4\sigma$, where $\sigma$ represents the smallest particle diameter. The interaction between particles $i$ and $j$ is the harmonic soft repulsive potential, 
\begin{equation}
V(r_{ij})=\left\{ \begin{array}{cr} \frac{\epsilon}{2}\left( 1-r_{ij}/\sigma_{ij}  \right)^2\quad & r_{ij}<\sigma_{ij} \\ 0 & r_{ij}\geq\sigma_{ij} \end{array} \right. ,
\end{equation}
where $r_{ij}$ is the distance between particle centers, $\sigma_{ij}$ is the sum of their radii, and $\epsilon$ sets the energy scale. We will take all particles to have equal mass $m$, measure energies in units of $\epsilon$, distances in units of the average particle diameter, and frequencies in units of $\sqrt{\epsilon/m \sigma^2}$. To obtain a jammed configuration at a target pressure, $p$, particles were initially placed at random in the simulation box with periodic boundary conditions (i.e. in an infinite temperature configuration). The system was then quenched to zero temperature by combining linesearch methods, Newton's method, and the FIRE algorithm \cite{quench}. The system was then incrementally expanded or compressed uniformly and then re-quenched to zero temperature until the target pressure was obtained to within $1\%$. For each configuration specified by a total number of particles of $256 \leq N \leq 10000$ and a pressure of $10^{-8}\leq p \leq 10^{-1},$ approximately 1000 states were prepared for analysis.

When using a purely repulsive potential there is a challenge in dealing with free surfaces; most notably, if one removes particles to create a surface in a finite-pressure jammed configuration, force balance would no longer be satisfied and the system would expand. We circumvent this problem by studying the corresponding ``unstressed'' network \cite{Alexander1998, Silbert2005pre}. We replace each pairwise interaction with a harmonic unstretched spring between nodes at the particle centers.  This gives us a system with the same geometry and connectivity as the original sphere packing. These unstressed networks are the cleanest way to understand the bulk density of states of the jammed particle packings, and can be used to understand, e.g., heat transport properties of the original system \cite{Vitelli2010}. They are also useful for systems with attractive interactions, such as Lennard-Jones systems \cite{Xu2007}.

We thus replace the jammed packing with the unstressed network. Formally, one constructs the $dN\times dN$ dynamical matrix $\mathcal{M}_{ij}$ by taking the second derivative of the energy: $\mathcal{M}_{ij} \equiv \frac{\partial^2 U}{\partial \vec{r}_i \partial \vec{r}_j}$, where 
\begin{equation}
U = \frac{1}{2} \sum_{\langle i,j \rangle} k_{ij} \left( (\vec{r}_i-\vec{r}_j) \cdot \hat{r}_{ij}\right)^2.
\end{equation}
Here $i$ and $j$ refer to particle indices, the sum is over neighboring particles, and $k_{ij} = \frac{\partial^2 V(r_{ij})}{\partial^2 r_{ij}}$ is the stiffness of the bond. Crucially, this expression for the dynamical matrix neglects terms proportional to stress that are present in the sphere packing. The pressure at which the sphere packing was prepared sets the average contact number for the unstressed system, and we thus use initial packing pressure as a proxy for the spring network connectivity. For the positive pressures and harmonic interactions considered in this work, the average excess contact number is $\langle Z \rangle-2d\sim p^{1/2}$ \cite{Ohern2003}. The dynamical matrix can be diagonalized to obtain the density of states, $D(\omega)$, of the unstressed spring network. In periodic jammed configurations the anomalous modes lead to a plateau in the a density of states that extends down to a characteristic frequency $\omega^*\sim \Delta Z$ \cite{Silbert2005}. Below this frequency, the density of anomalous modes drops to zero. In the following, we will report measurements with respect to an estimate of $\omega^* \approx 2 \sqrt{p}$, which is approximately the frequency at which the density of states for bulk systems drops below 1.

With the unstressed spring network in hand, we create a free surface by removing any bond that crosses a boundary of interest. In this work we focus on systems in a thin film or slab geometry, and so remove the periodic boundary conditions in the $x$-direction. This is equivalent to cutting any bond that crosses $x=0$ or $x=L$ where $L$ is the linear system size.  Our system is thus a strip of width $L$ in the $x$-direction, with periodic boundary conditions in the remaining directions.

\section{Numerical results \label{sec:harmresults}}

\subsection{Density of States \label{sec:dos}}
\begin{figure}[htb!]
\centering{\includegraphics[width=0.9\linewidth]{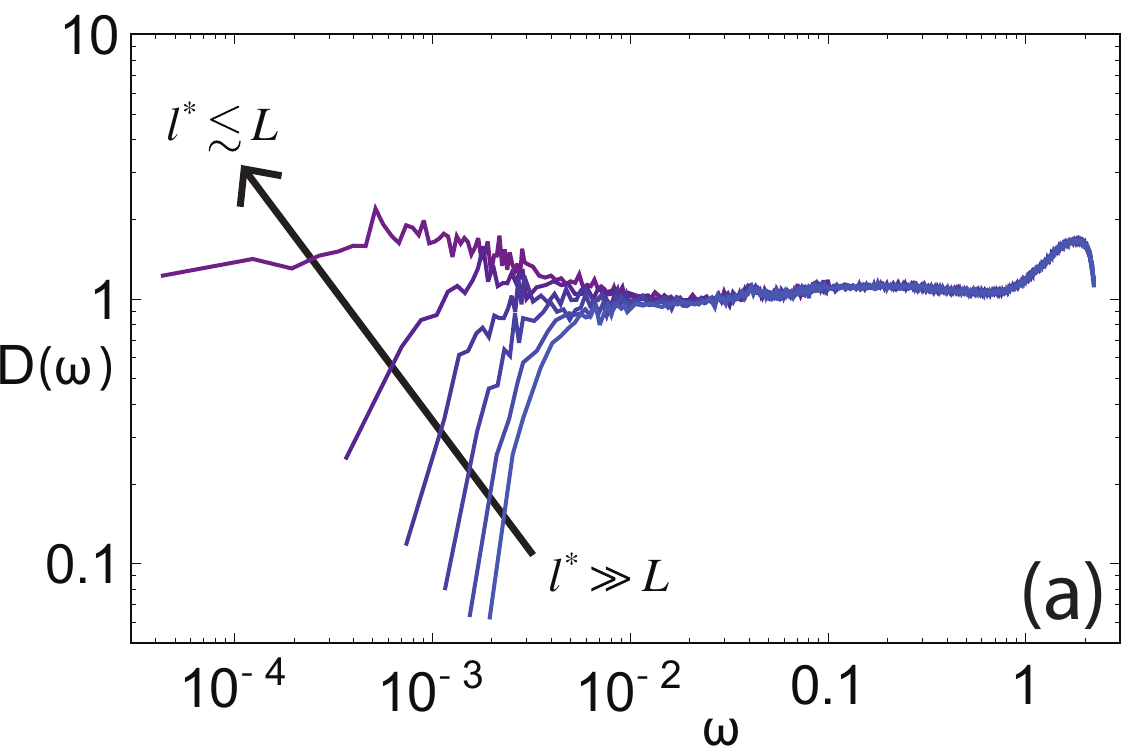}}
\centering{\includegraphics[width=0.9\linewidth]{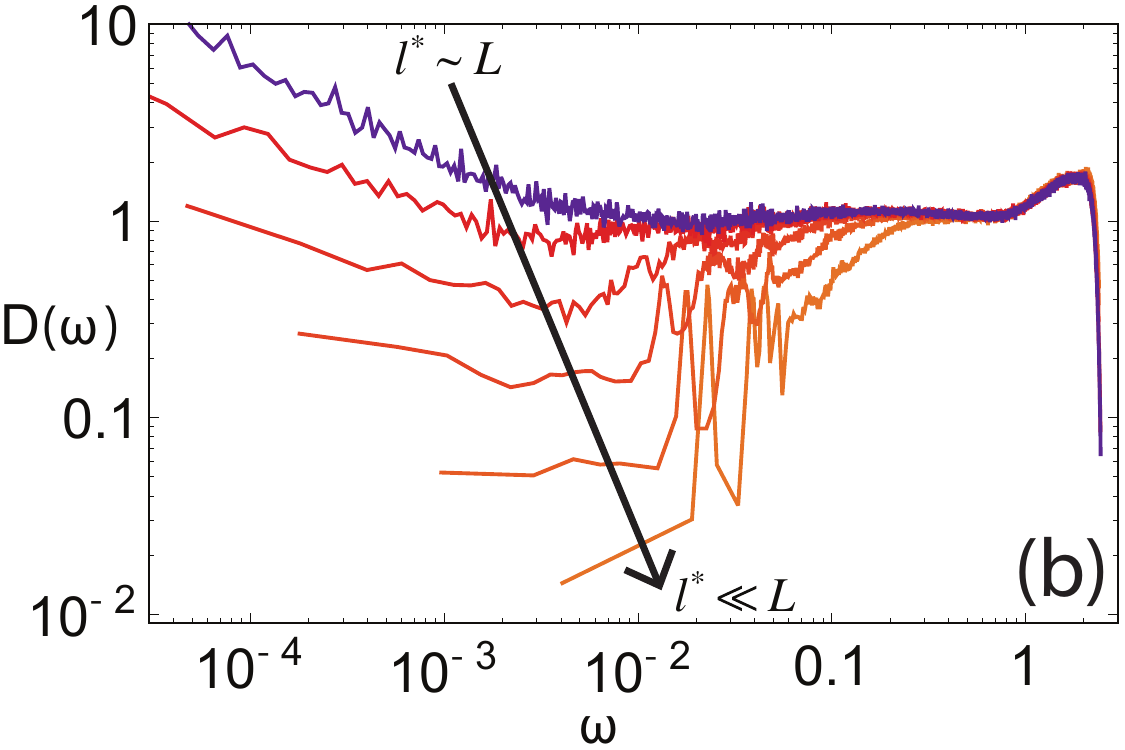}}
\centering{\includegraphics[width=0.9\linewidth]{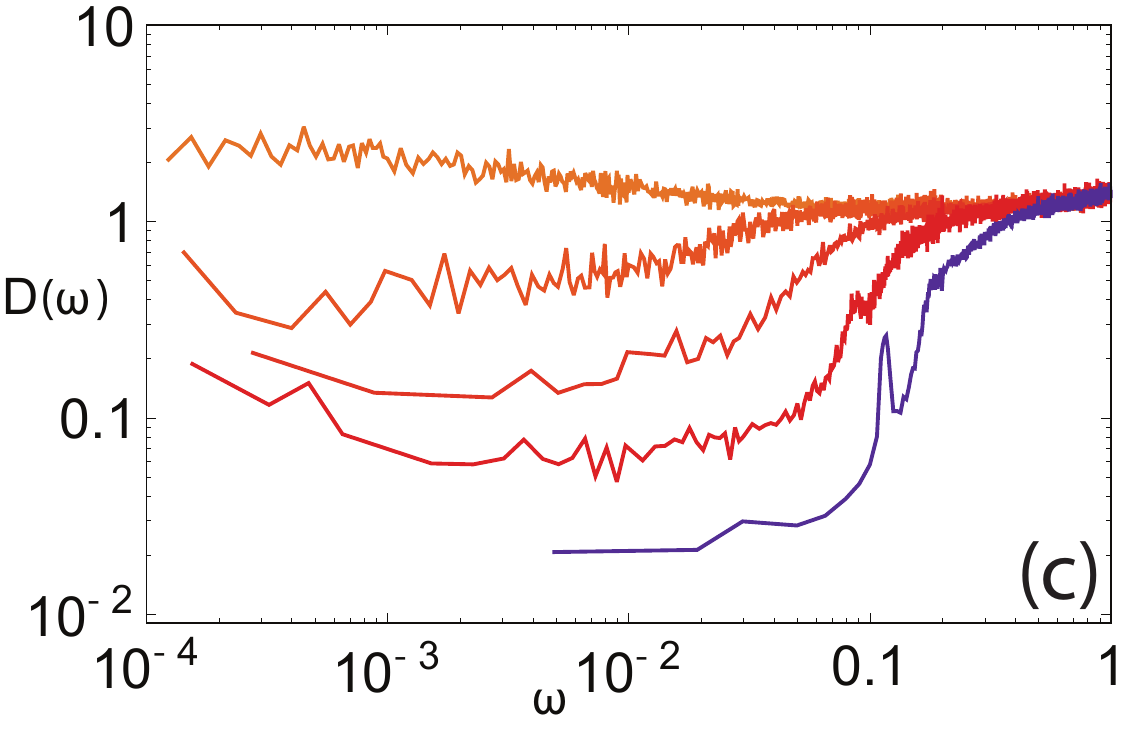}}
\caption{\label{fig:dos}
(a) and (b) Density of vibrational modes averaged over 2500 networks derived from  jammed packings of $N=500$ particles in two dimensions. (a) From right to left the pressures of the initial packings are $p=4.0\times 10^{-6},\ 6.3\times 10^{-6},\ 1.0\times 10^{-5},\ 1.6\times 10^{-5},\ 2.5\times 10^{-5},$ and $4.0\times 10^{-5}$, for which $L \lesssim l^*$. (b) From top to bottom the pressures of the initial packings are $p=1.0\times 10^{-4},\ 2.5\times 10^{-4},\ 6.3\times 10^{-4},\ 1.6\times 10^{-3},\ 4.0\times 10^{-3},$ and $1.0\times 10^{-2}$, for which $L \gtrsim l^*$. (c) Low-frequency part of the density of vibrational modes for systems of $N=1000$ particles in 3D. From top to bottom the pressures of the initial packings are $p=6.3\times 10^{-4},\ 1.6\times 10^{-3},\ 4.0\times 10^{-3}, \ 1.0\times 10^{-2},$ and $2.53\times 10^{-2}$, for which $L \gtrsim l^*$.}
\end{figure}

We begin by characterizing the density of vibrational modes in these free-surface systems. Figure \ref{fig:dos} shows representative examples of the density of states that we obtain by cutting free surfaces at $x=0$ and $x=L$ in both 2 and 3 dimensions. The different curves correspond to different pressures at which the harmonic disk packings were originally prepared. As noted above, before cutting the free surface the pressure sets the characteristic length scale $l^*\sim\Delta Z^{-1}$, and by varying the initial pressure of the packings we are able to study the density of states as a function of the ratio $l^*/L$. Although it may be more intuitive to study this ratio by varying the system size, in practice it is much easier to prepare systems at a fixed size and minimize them to different targeted pressures. We note in passing that at all values of $l^*/L$ that we study our disordered packings have of order $L^{d-1}$ surface zero-frequency modes~\cite{Thorpe1995}: for modestly over-constrained systems there is a $\Delta Z$-dependent, non-zero probability per unit surface area of creating a localized zero-frequency mode, and the resulting modes are localized to the surface on the scale of the particle size \cite{Goodrich2013}. In addition to these zero frequency modes, however, there is also a nontrivial population of finite-frequency modes associated with the free surface.

When the strip thickness is $L=l^*$, the system as a whole is brought very close to the isostatic threshold and, by analogy with bulk systems \cite{Silbert2005}, one expects a plateau in the density of states extending to arbitrarily low frequencies. When the strip thickness is $L<l^*$, the system is brought below isostaticity by the introduction of free surfaces and is no longer rigid. For finite-sized systems the lowest-frequency plane wave has a frequency proportional to $1/L$, and there are no disordered modes in the frequency range $0 < \omega \lesssim \omega^*$ \cite{Wyart2013}. This leads to an effective gap in the density of vibrational modes, as seen in Fig. \ref{fig:dos}a. The figure shows a larger gap at lower initial packing pressures, corresponding to a larger ratio of $l^*/L$. Not shown is the delta-function spike of additional \emph{extended} zero-frequency modes that grows as the system is taken farther and farther below the isostatic point by increasing $l^*/L$.

Our primary focus is on systems with $l^*/L < 1$.  These systems have free surfaces but remain rigid because the system retains enough contacts to be globally stable. Just as in the periodic case, there is a plateau that extends down to a frequency $\omega^*_s$. We find that for strips this frequency is a factor of two smaller than the lower frequency edge of the plateau in identical systems with full periodic boundary conditions, $\omega^*_s \approx \omega^*/2$. This result is consistent with a cutting argument, as we will show in the Discussion.   The most noticeable feature of Fig.~\ref{fig:dos}b, however, is a secondary population of modes below $\omega^*_s$ that is absent in the periodic system. This feature persists for three-dimensional systems with cut surfaces, as shown in Fig.~\ref{fig:dos}c. The additional modes appear to extend all the way down to zero frequency; the curves end at low frequencies where we no longer have sufficient statistics.  Note that for each $l^*/L< 1$ there is an upturn at very low frequencies.  This upturn is particularly striking at $l^*/L \sim 1$.  An extremely minor upturn has been observed for periodic jammed systems with $\Delta Z < 3\times 10^{-2}$ \cite{Silbert2005pre}, but here we see an apparent power-law increase in the density of states that scales as $\omega^{-1/2}$ at low frequencies. This feature has not been understood in the context of the counting/variational argument \cite{Wyart2005}, and is currently unexplained. 

\begin{figure}[htb!]
\centering{\includegraphics[width=0.9\linewidth]{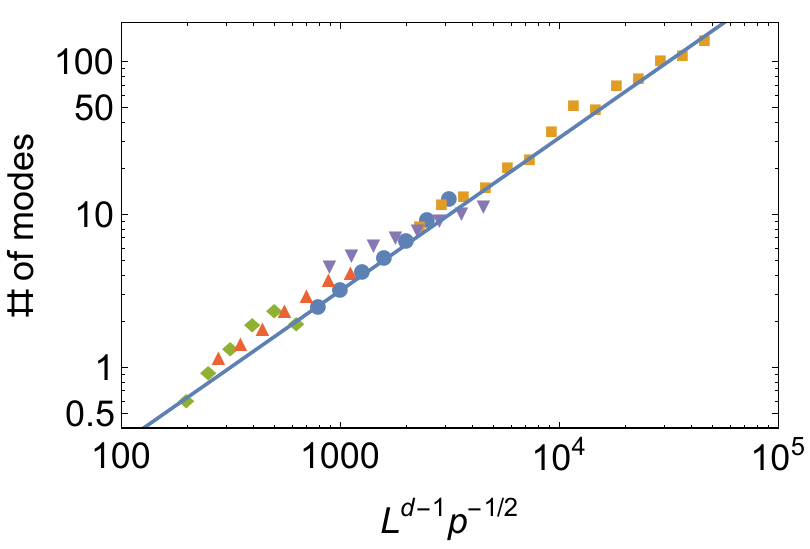} }
\caption{\label{fig:doscollapse}
Log-log plot of the number of modes below $\omega^*_s/2$ versus $L^{d-1}p^{-1/2}$. Points are drawn from two-dimensional packings with $N=250,\ 500,\ 2048$ and three-dimensional packings with $N=1000,\ 10000$. The straight line is a guide to the eye with unit slope.}
\end{figure}

The number of modes in this secondary, low-frequency portion of the density of states strongly suggests that this contribution to the density of states arises from the existence of free surfaces. To ensure that we do not include modes that are present in the bulk, we count only the number of modes below $\omega^*_s/2$. For different system sizes the average number of modes per system in the frequency range $0<\omega <\omega^*_s/2$ scales with the free surface area, $L^{d-1}$, as expected. Additionally, at fixed system size but with varying initial packing pressure we find that the number of modes in this frequency range per system scales as $1/\sqrt{p}\sim (l^*)$. These two features are shown in Fig. \ref{fig:doscollapse}, which plots the number of low-frequency modes versus $L^{d-1} p^{-1/2} $ for a variety of pressures and system sizes in both two and three dimensions. This scaling suggests that the volume of particles that participate in surface modes with $0 < \omega < \omega^*_s/2$ scales as $L^{d-1} l^*$; assuming that surface modes are localized to the surface leads to the conclusion that particles within $l^*$ of the free surface participate in these modes.

\subsection{Surface mode structure \label{sec:struct}}

\begin{figure}[htb!]
\centering{\includegraphics[width=0.9\linewidth]{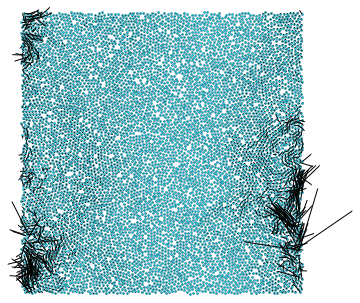}}
\centering{\includegraphics[width=0.9\linewidth]{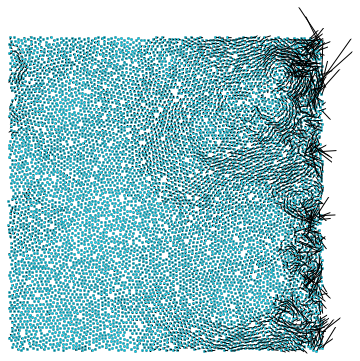} }
\caption{\label{fig:mode}
Typical low-frequency modes for two-dimensional systems with periodic boundary conditions along the top and bottom edges of the cell and free boundaries along the vertical edges. Circles represent particle centers and black lines represent the orientation and magnitude of particle motion, $\delta\vec{R}_{i}$, in that mode. The frequencies correspond to $\omega/\omega^*_s=0.24$ (top) and $\omega/\omega^*_s=0.62$ (bottom).}
\end{figure}

We can now look at the spatial structure of the modes that lie in the new band between $\omega=0$ and $\omega=\omega^*_s$ .  Figure \ref{fig:mode} shows two typical examples of these modes in a two-dimensional system. The black lines show the magnitude and orientation of the polarization vector of the given mode on each particle. The modes are clearly localized to the free surface. As seen in the left figure, we occasionally find modes that tunnel through the sample and have localized vibrations at both free surfaces. Additionally, we typically find that the extent of localization is weakly frequency-dependent, with a localization length that grows with frequency. A quantification of this dependence is difficult, though, as individual modes typically have non-trivial structure, including plane-wave contributions. 

\begin{figure}[htb!]
\centering{\includegraphics[width=0.9\linewidth]{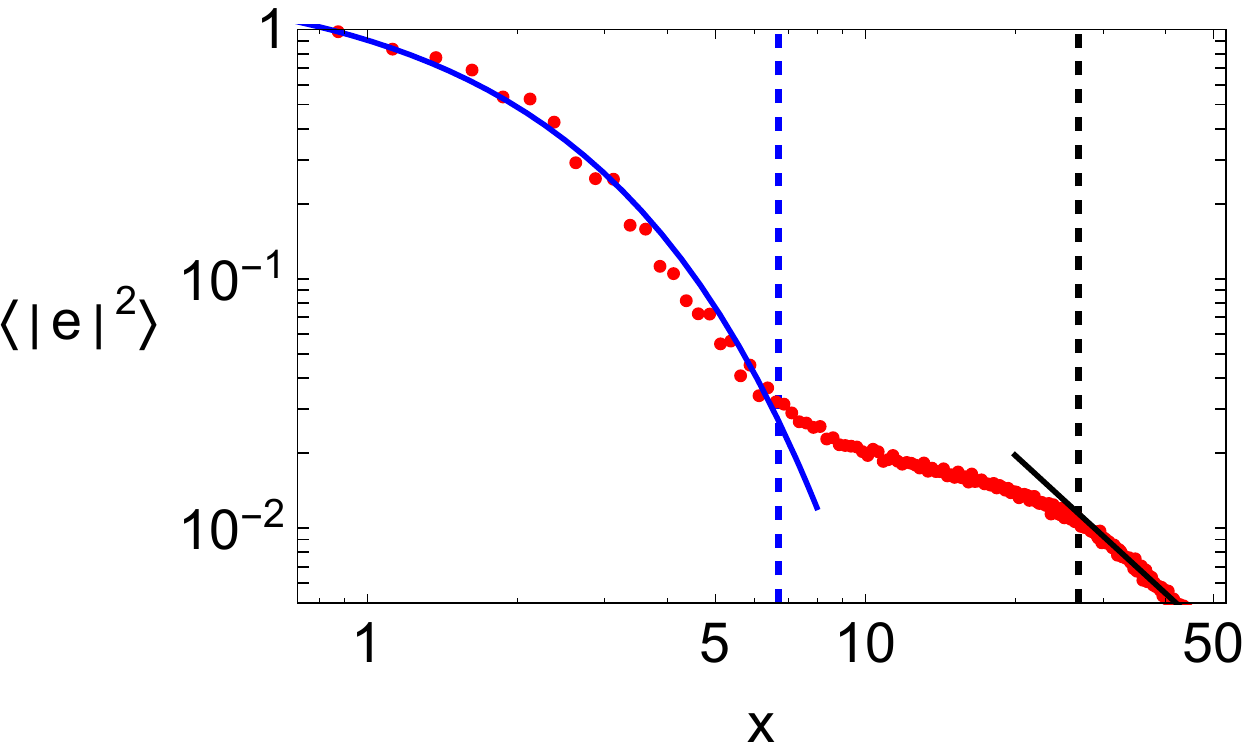}}
\centering{\includegraphics[width=0.9\linewidth]{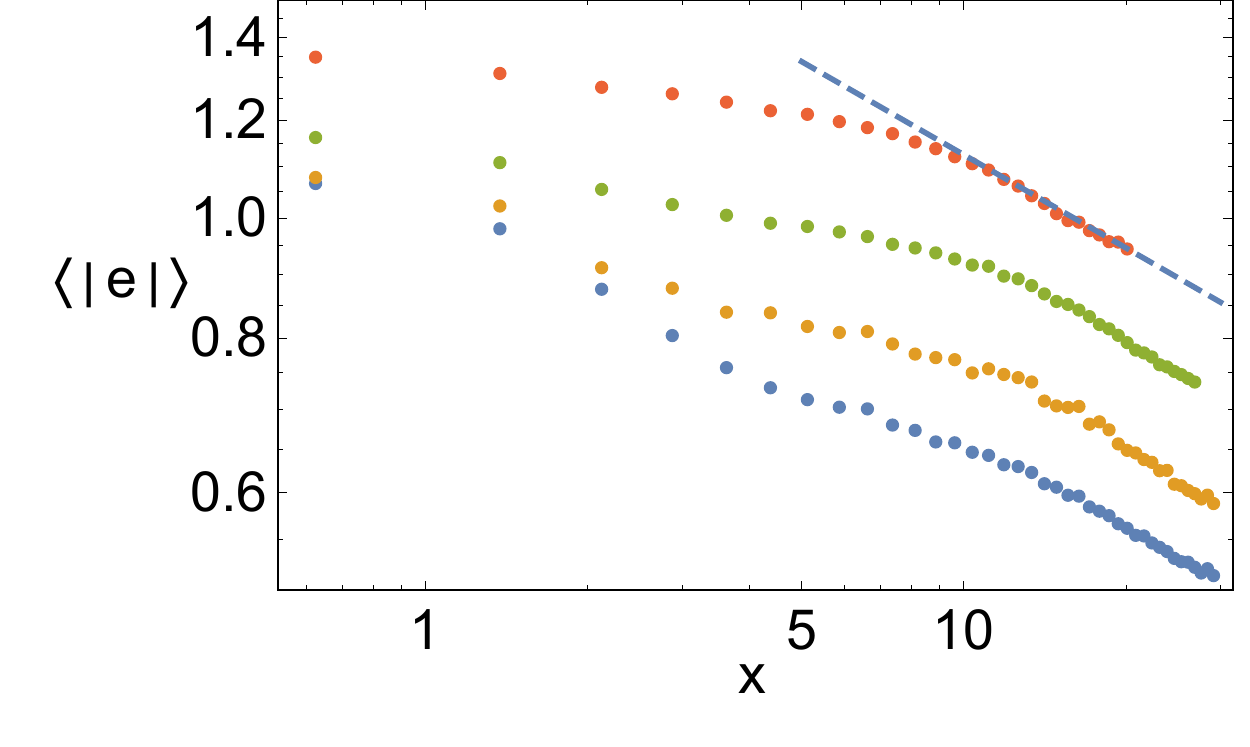}}
\caption{\label{fig:overlap}
(top) Log-log plot of the overlap function for two-dimensional packings with $N=10000$ and $p=1.0\times 10^{-3}$, vertically shifted (normalized by $\langle |e|^2 \rangle$ of the leftmost data point) for clarity. The solid blue line is an exponential fit over the first region from the surface to the blue dashed line, while the solid black line is a straight line fit on the log-log plot that characterizes the third region (from the black dashed line to the center of the sample). The vertical dashed lines show where the data deviate by a fixed percentage from the solid fitting lines, with the black vertical dashed line marking a knee separating the second from the third regimes. (bottom) Log-log plot of the polarization magnitude between $x$ and $dx$ for $N=10000$ and pressures of (bottom to top $p=4.0\times10^{-3},\ 6.3\times 10^{-3}, 1.0\times 10^{-2},\ 1.6\times 10^{-2}$. The dashed line is a guide to the eye with slope $-1/4$. The curves have been shifted vertically for clarity.}
\end{figure}

In order to quantify the decay of the vibrational amplitude from the surface into the bulk, we average the vibrational amplitude over all modes in the frequency band $0 < \omega \leq \omega^*_s$.  Specifically, we look at the average polarization magnitude and average squared polarization magnitude of particles between $x$ and $dx$ as a function of distance, $x$, from the free surface (similar to the overlap function defined by Wyart \cite{Wyart2005annrev}):
\begin{equation}\label{eq:overlap}
\langle | \vec{e}|^\tau \rangle dx = \sum_{\mu}  \sum_{x_i\in [x,x+dx]} \left| \delta\vec{R}_{i,\mu} \right|^\tau.
\end{equation}
Here $\mu$ indexes any of the modes whose frequency is in the surface plateau region, $\delta\vec{R}_{i,\mu}$ refers to the vector displacement of particle $i$ in vibrational mode $\mu$, and $\tau=1$ or $\tau=2$. We have checked that our subsequent results are insensitive to the choice of upper frequency cut-off in the set of modes we study, as long as that cut-off is less than $\omega^*_s$. A representative plot of this surface-mode profile is shown in Fig. \ref{fig:overlap}.

\begin{figure}[htb!]
\centering{\includegraphics[width=0.9\linewidth]{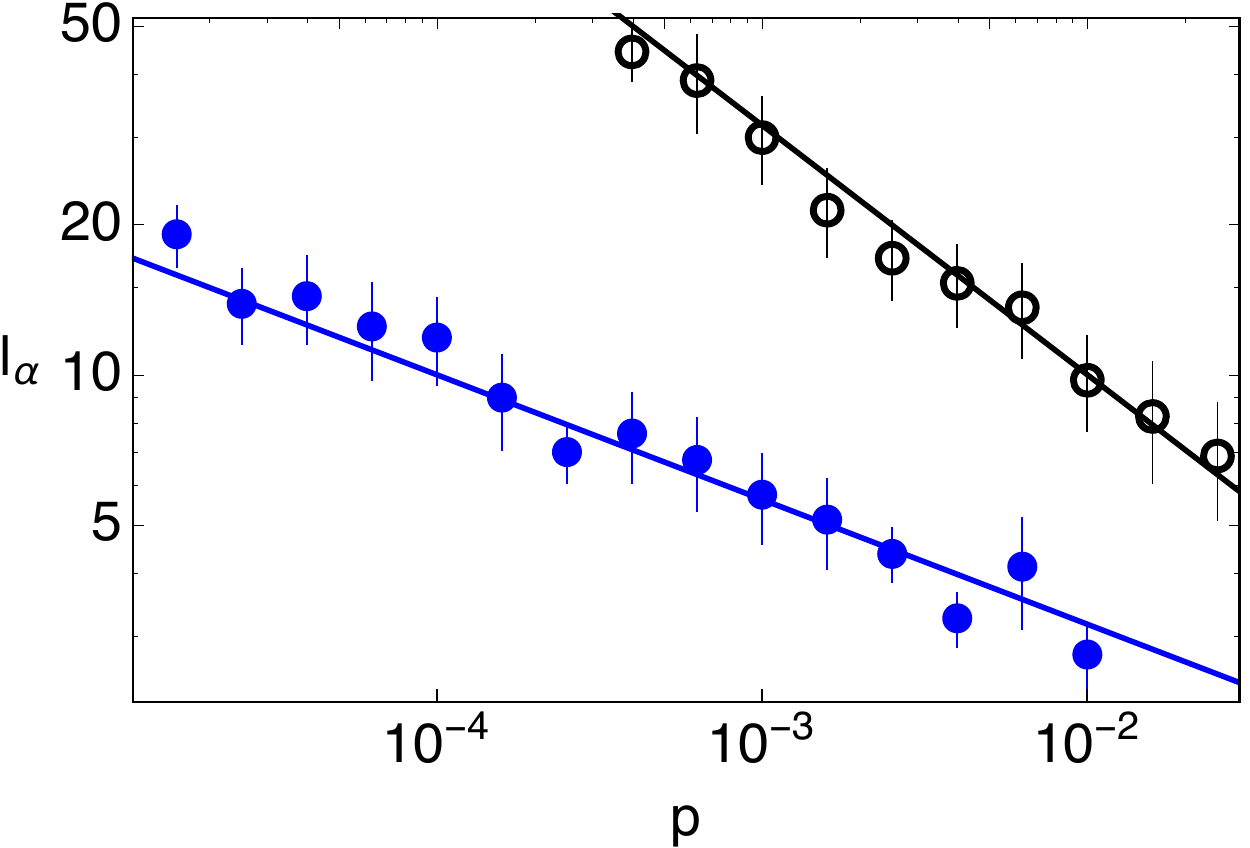} }
\caption{\label{fig:pdepth}
Length scales marking the end of the exponential decay regime at small $x$ corresponding to the blue dashed line in Fig.~\ref{fig:overlap}  (blue solid circles) and the crossover between the second and third regimes corresponding to the black dashed line in Fig.~\ref{fig:overlap}  (black open circles), as a function of pressure for $N=10000$ particle systems in 2D. Straight lines are guides to the eye with slopes $-0.25$ and $-0.50$, respectively; these pressure-dependancies correspond to the scalings of $l_T$ and $l^*$. The error bars around each point correspond to the range of values obtained by varying the parameters of the fitting procedure described in the main text.}
\end{figure}

As shown in Fig. \ref{fig:overlap}(a), the average mode profile decreases away from the surface. The blue curve, an exponential decay, is a good fit to the region closest to the surface.  The profile begins to deviate from the initial exponential decay at a distance that we mark in Fig.~\ref{fig:overlap}(a) with a vertical blue dashed line. We have studied mode profiles as a function of initial pressure, and for sufficiently low pressures we consistently see that close to the surface the profile has a clear exponential decay, and that the distance over which this exponential decay persists decreases with increasing pressure. At the highest pressures studied, when an extrapolation would suggest that the exponential decay length is less than the $\approx 2\sigma$ length scale over which the jammed packings have a non-trivial local structure, it is harder to observe this exponential decay.  We have also confirmed that the same length scale can be obtained by fitting an exponential decay on a mode-by-mode basis, although this leads to a much noisier signal. In Fig.~\ref{fig:pdepth} we plot (blue solid circles) the distance at which the average mode profile deviates from an exponential decay, corresponding to the blue dashed line in Fig.~\ref{fig:overlap}, as a function of pressure.  We find that this distance scales as the transverse length scale, $l_T\sim p^{-1/4}$, which diverges at the jamming transition~\cite{Silbert2005,Schoenholz2013,Lerner2014}. By varying the precise region over which we fit and the tolerance at which we declare the profile to have deviated from the fit we obtain the error bars in Fig. \ref{fig:pdepth}. 

That the modes decay on the scale of the transverse length is surprising in light of our analysis of the density of states, where we found of order $\sim L^{d-1}l^*$ modes below $\omega^*_s/2$. Since $l^* > l_T$, this suggests that even though the dominant decay length is on the scale of $l_T$, there must be contributions from particles farther away from the surface, i.e. on the scale of $l^*$.  In fact there are indications of this length scale in the surface mode profile.  Although the average mode structure beyond $l_T$ is complicated  by the finite number of plane waves that may lie in the frequency band $0<\omega<\omega^*$, we find that the initial exponential decay is consistently followed by a crossover regime which ends with a knee. At larger $x$ the decay is again faster, indicating a new regime. The onset of this new regime is marked by a vertical black dashed line in Fig.~\ref{fig:overlap}(a). Although we have a very limited range in this third regime, the decay in this regime has the same slope on a log-log plot across the range of pressures for which the third regime is observable in our $N=10000$ two-dimensional systems.  This is shown in Fig.~\ref{fig:overlap}(b), where we plot the mode profiles on a log-log plot for several pressures, with vertical shifts, to show that they have the same slope in this third regime.

We plot the distance corresponding to the onset of the third regime as a function of pressure in Fig.~\ref{fig:pdepth} (black open circles).  We find that the onset of the third regime of decay scales as $l^*\sim p^{-1/2}$, which diverges at the jamming transition~\cite{Silbert2005,Wyart2005,Goodrich2013,Ellenbroek2006}.  This is consistent with our expectation that, based on the scaling of the surface density of states, these surface modes should extend into the system on the length scale $l^*$.  

In summary, the surface modes appear to have a signature of both of the two diverging length scales associated with jamming~\cite{Silbert2005, Wyart2005, Goodrich2013, Schoenholz2013, Ellenbroek2006, Lerner2014}.  The 10000-particle systems studied have a box size of roughly $100\sigma \times 100\sigma$, which accounts for our inability to observe $l^*$ at very low pressures: when the second regime of the mode profile extends past $\sim 50\sigma$ it cannot be reliably detected as the second free surface starts influencing the decay of the overlap function. Thus, studying the transition between the secondary and tertiary decay regimes for lower pressures would require much larger systems. Additionally, as noted above there is local structure on a scale of $\sim 2\sigma$, and so when $l_T$ is comparable to this distance (at very high pressures) it, too, cannot be reliably observed.

\section{Discussion\label{sec:bridge}}

The most striking feature of the density of states for strips of finite width is the presence of a new population of disordered surface modes with frequencies below $\omega^*_s$.  We identify these as surface modes because their number scales as $L^{d-1}$.  Above $\omega^*_s$, there is a plateau in the density of states that scales with the volume of the strip, $L^d$.  The frequency $\omega^*_s$ is half that of the frequency $\omega^*$, which marks the lower frequency edge of the plateau in the bulk density of states.  This factor of two may be understood in the context of a simple counting estimate for $l^*$ and $\omega^*$ \cite{Wyart2005, Goodrich2013}. In bulk systems a counting estimate of $l^*$ comes from thinking about cutting a boundary completely around the system on a size scale $L$. The number of bonds cut by this procedure is $N_c^{cut}=\gamma \langle Z \rangle L^{d-1}$, and the number of excess bonds (above isostaticity) the system had before the cut is $N_c^{extra} = \nu \Delta Z L^d$, where $\gamma$ and $\nu$ are prefactors that depend on the geometry of the cut. Estimating $l^*$ as the length at which $N_c^{cut}=N_c^{extra}$ yields $l^*\sim \gamma \langle Z \rangle/(\nu \Delta Z)$. However, in a system that already has free surfaces in one of the dimensions there is a reduction in $N_c^{extra}$ by a surface term: $N_c^{extra} =  \nu \Delta Z L^d - \gamma \langle Z \rangle L^{d-1}$. Equating $N_c^{extra}$ and $N_c^{cut}$ for these free-surface systems thus increases the counting estimate of $l^*$ by a factor of two, and hence $\omega^*\sim1/l^*$ is reduced by a factor of two.

A more pressing question to address is why the surface modes fill in the gap $0 < \omega \leq \omega^*_s$, with a number of modes in this regime that scales as $l^* L^{d-1}$. The fact that the surface modes can have arbitrarily low frequencies is a consequence of the arguments of Goodrich et al. \cite{Goodrich2014pre}, where it was noted that, depending on the degree of localization of a given mode, breaking a contact can lower that individual mode's frequency by an arbitrary amount. Thus, if we assume that modes are quasi-localized to the surface, cutting $L^{d-1}$ bonds could generically create a population proportional to $\sim L^{d-1}$ of very-low energy modes (since this is related to the probability of cutting a bond important to one of those quasi-localized modes). The scaling of the size of this population of sub-$\omega^*_s$ modes is independent of the geometry of the cutting, but the actual number of such modes and their distribution in frequency could depend on the spatial distribution of cut bonds. In the case of a surface, then, why does the surface density of states scale as $l^*$?

A justification comes from recalling that if the system has $L < l^*$ then it loses its rigidity~\cite{Wyart2005, Goodrich2013}.  One reasonable assumption is that this rigidity loss occurs because very soft surface modes that decay from each cut surface to a distance $l^*$ can communicate with each other through the system once $L\sim l^* $. We observe that there are two decay lengths governing the decay of the surface mode profile, $l_T$ and $l^*>l_T$.  If surface vibrations are localized on a scale of $l^*$, one expects, from a straightforward generalization of the variational argument of Wyart et al. \cite{Wyart2005}, that some population of them (of order $L^{d-1}$) would have an energy cost bounded by $\delta E_{loc} \lesssim (l^*)^{-2}$ and thus have a frequency $\omega \lesssim \omega^*$. (This does not preclude the possibility of ``surface'' modes additionally appearing at higher frequencies.)  Thus, the assumption that modes are localized to be within $l^*$ of the surface -- which is verified by the appearance of $l^*$ as a decay length in the surface mode profile -- immediately suggests a population of $l^*L^{d-1}$ modes at frequencies below $\omega^*_s$, consistent with our observation.  

We note that our observation of two decay lengths in the profile, $l_T$ and $l^*$, is consistent with ideas of Lerner et al. \cite{Lerner2014}, which suggests that  $l_T$ is the length scale below which disordered response, beyond that predicted by continuum elasticity, can be observed as long as the system is at least $l^*$ in size. 

It is natural to ask what we might expect for surface modes in disordered systems with longer-range interactions. We speculate that our findings may have implications for the existence of a free-surface length scale in Lennard-Jones thin films. Although these systems do not properly have a jamming transition (it lies inside the liquid-vapor spinodal \cite{Berthier2009}), and the surface modes share the same frequency range as bulk vibrational modes \cite{Jain2004}, there may still be a remnant of the two surface length scales seen in our present studies. One can define longitudinal and transverse length scales by comparing the speeds of sound with the boson peak frequency. For instance, $l_T\sim c_T/\omega^*$, where the transverse speed of sound is $c_T=\sqrt{G/\rho}$, with $G$ the shear modulus and $\rho$ the mass density. In jammed systems this definition recovers the expected scalings of $l_T\sim p^{-1/4}$ and   $l^*\sim p^{-1/2}$~\cite{Silbert2005}. We can estimate these length scales by estimating the boson peak and moduli of a zero-temperature Lennard-Jones glass whose density corresponds to a zero-pressure state. Doing so, we find that $l_T\sim2.5\sigma$ and $l^*\sim 6.0\sigma$. While modest, the estimated $l^*$ is longer than static length scales typically observed near free surfaces, and is roughly consistent with the characteristic size of the mobile layer of Lennard-Jones polymer glasses below their glass transition.  It is therefore possible that the length scales we observe for jammed systems may survive in the surface properties of real glassy thin films.

\begin{acknowledgments}
We thank Samuel S. Schoenholz and Randall D. Kamien for useful discussions. This research was supported by the U.S. Department of Energy, Office of Basic Energy Sciences, Division of Materials Sciences and Engineering under award DE-FG02-05ER46199 (DMS, CPG, and AJL) and  DE-FG02-03ER46088 (DMS and SRN), the UPENN MRSEC under award NSF-DMR-1120901 (DMS and AJL), and the Advanced Materials Fellowship of the American Philosophical Society (DMS).
\end{acknowledgments}

\bibliography{Surface_modes_bib}

\end{document}